\documentclass[aps,prd,showpacs,twocolumn,standalone,10pt,superscriptaddress,floatfix,nofootinbib,longbibliography]{revtex4-2}
\usepackage[T1]{fontenc}
\usepackage[utf8]{inputenc}
\usepackage{amssymb,amsfonts,amsmath,bm}
\usepackage{graphicx}
\usepackage{dcolumn}
\newcommand{\fourrowlabel}[1]{\smash{\raisebox{-1.5\normalbaselineskip}{#1}}}
\usepackage[
colorlinks=true,
linkcolor=blue,
breaklinks=true,
urlcolor=blue,
citecolor=blue]{hyperref}

\begin{document}

\makeatletter
\def\bibsection{%
  \par
  \section*{\refname}%
  \@nobreaktrue
}
\makeatother

\title{GPU-Accelerated Matrix-Based Hough Transform for Online Track Reconstruction in the STCF MDC}

\author{Liang Peng}\thanks{These authors contributed equally to this work.}
\affiliation{School of Physics and Electronic Science, Hunan University of Science and Technology, Xiangtan 411201, China}
\affiliation{Hunan Provincial Key Laboratory of Intelligent Sensors and New Sensor Materials, School of Physics and Electronic Science, Hunan University of Science and Technology, Xiangtan 411201, China}
\author{Baolin Zhang}\thanks{These authors contributed equally to this work.}
\affiliation{School of Physics and Electronic Science, Hunan University of Science and Technology, Xiangtan 411201, China}
\author{Zhaoli Guo}
\affiliation{School of Physics and Electronic Science, Hunan University of Science and Technology, Xiangtan 411201, China}
\author{Jiarui Zhao}
\affiliation{School of Physics and Electronic Science, Hunan University of Science and Technology, Xiangtan 411201, China}
\author{Yizhen Ma}
\affiliation{School of Physics and Electronic Science, Hunan University of Science and Technology, Xiangtan 411201, China}
\author{Aonan Zhu}
\affiliation{School of Physics and Electronic Science, Hunan University of Science and Technology, Xiangtan 411201, China}
\affiliation{Hunan Provincial Key Laboratory of Intelligent Sensors and New Sensor Materials, School of Physics and Electronic Science, Hunan University of Science and Technology, Xiangtan 411201, China}
\author{Huilin Li}\email[First corresponding author:~]{lihuilin@hnust.edu.cn}
\affiliation{School of Physics and Electronic Science, Hunan University of Science and Technology, Xiangtan 411201, China}
\affiliation{Hunan Provincial Key Laboratory of Intelligent Sensors and New Sensor Materials, School of Physics and Electronic Science, Hunan University of Science and Technology, Xiangtan 411201, China}
\author{Zhujun Fang}\email[Second corresponding author:~]{fzj19@ustc.edu.cn}
\affiliation{Department of Modern Physics, School of Physical Sciences, University of Science and Technology of China, Hefei 230026, China}
\affiliation{Laboratory of Nuclear Detection and Nuclear Electronics, University of Science and Technology of China, Hefei 230026, China}

\date{\today}

\begin{abstract}
The Super Tau-Charm Facility (STCF) is a proposed next-generation high-luminosity electron--positron collider operating at center-of-mass energies of 2--7 GeV for precision studies of tau--charm physics. Its high event rate, detector occupancy, and background level impose stringent requirements on real-time track reconstruction in MDC, particularly for low-transverse-momentum particles with strongly curved or multi-turn trajectories. To address this challenge, we develop a GPU-accelerated matrix-based Hough transform method for online track reconstruction in the STCF MDC. Following an algorithm--architecture co-design paradigm, the data representation and computational workflow of the conformal Hough transform are reformulated for GPU execution. The original irregular parameter-space computations are organized into regular matrix-based operations, and the core computations are adapted to CUDA thread organization and the GPU memory hierarchy to exploit the inherent parallelism of the Hough transform and reduce computational and data-transfer overhead. Tests on five representative simulated physics channels with nominal background overlay show an average signal retention ratio of 93.04\%, while reducing the retained hit volume to 34.92\% of the original level. The GPU implementation processes 1,000 events in approximately 0.14 s and achieves a maximum speedup of 151.57$\times$ over a CPU implementation following the same algorithmic workflow. These results demonstrate that the proposed method substantially improves track reconstruction throughput while preserving track-associated hits, providing a new methodological perspective for real-time track reconstruction in future high-luminosity particle-collider experiments.
\end{abstract}

\maketitle

\section{Introduction}\label{sec:introduction}

The Super Tau-Charm Facility (STCF) is a next-generation high-luminosity electron--positron collider facility currently under development in China\cite{ref1,ref2}. It is designed to conduct precision measurements and searches for new physics in the center-of-mass energy range of 2--7 GeV. A peak luminosity exceeding $0.5\times10^{35}~\mathrm{cm}^{-2}\mathrm{s}^{-1}$ is planned at a center-of-mass energy near 4 GeV. High-luminosity operation will substantially increase the event rate, while also increasing the complexity of beam-related backgrounds and the volume of detector data\cite{ref3}. These conditions impose more stringent requirements on data transmission, online computing, and long-term storage. Without effective online filtering and data compression, the continuously increasing volume of raw data would rapidly increase the burden on the computing and storage systems. Therefore, the online trigger system must not only select physics events but also provide reliable fast-reconstruction results under stringent latency constraints. These results provide essential information for subsequent event selection, data compression, and offline analysis\cite{ref4}.

The high-level trigger (HLT) system is a key component of online data processing in future high-energy physics experiments. Track reconstruction is one of its most fundamental and computationally intensive tasks\cite{ref5,ref6}. It reconstructs charged-particle trajectories from discrete hits recorded by subdetectors, such as main drift chambers and silicon detectors. It also estimates the transverse momentum, azimuthal angle, and vertex-related parameters of each track. The track-reconstruction results directly affect the event-selection efficiency, background-rejection capability, and the quality of subsequent particle identification and physics analysis. Therefore, track reconstruction algorithms for the STCF HLT must simultaneously provide high reconstruction efficiency, strong robustness against backgrounds, low latency, and high throughput.

The Main Drift Chamber (MDC) is an important subdetector for measuring charged-particle trajectories \cite{ref7}. It also contributes to combined triggering across multiple subdetectors and to event selection. Under high-luminosity operating conditions, finite detector spatial resolution, multiple scattering in detector materials, electronic noise, and beam-related backgrounds can cause measured hits to deviate from ideal trajectories. These effects can also lead to locally missing hits, spurious hits, and spatial overlap between different tracks. Low-transverse-momentum tracks are more strongly curved and may have shorter effective arc lengths. They may also traverse fewer detector layers and provide fewer usable hits. These characteristics further increase the difficulty of candidate-track identification and parameter estimation. Such effects not only reduce the stability of track-parameter estimation but also enlarge the combinatorial search space. Conventional offline reconstruction relies on iterative combinatorial searches and precision fitting\cite{ref8}. Although it can achieve high reconstruction accuracy, it cannot be directly applied to the HLT because of its stringent latency and throughput requirements. Online track reconstruction therefore requires a candidate-track search method that can rapidly identify geometrically consistent hit patterns, remain robust against noise and missing hits, and efficiently exploit parallel computing architectures.

The parameter-space voting mechanism of the Hough transform provides a viable approach to addressing these challenges. It maps discrete points from the original space into a parameter space. Points belonging to the same geometric object then vote for identical or neighboring parameter cells. A local peak in the accumulator can consequently be interpreted as a candidate geometric object\cite{ref9}. Because votes from multiple points aggregate global support, the Hough transform retains a certain degree of robustness in the presence of noise points, locally missing data, or incomplete edges. Compared with candidate-by-candidate combinatorial searches, the voting mechanism does not require all possible hit combinations to be enumerated in advance. It is therefore more suitable for generating track candidates under high-occupancy conditions. Previous studies have applied the Hough transform and its multidimensional extensions to track recognition in complex detector environments\cite{ref10,ref11,ref12}. Their potential for fast pattern recognition and online triggering has also been investigated in the context of the LHC and its upgrades\cite{ref13}. Randomized, probabilistic, hierarchical, and parallel variants developed in image processing further demonstrate that the Hough transform offers clear geometric interpretability, relatively high noise tolerance, and a voting procedure amenable to parallelization\cite{ref14,ref15,ref16,ref17,ref18}.

However, parameter-space voting in the Hough transform incurs substantial computational and memory overhead\cite{ref19,ref20,ref21}. Its deployment in the HLT requires the algorithm to be restructured in terms of its mathematical formulation, data organization, and parallel execution model\cite{ref22}. Such restructuring is necessary to alleviate computational and memory-access bottlenecks\cite{ref23,ref24}. If serial traversal and random memory access are retained, contention during accumulator updates and the cost of peak finding will further limit online throughput. Graphics processing units (GPUs) provide massive thread-level parallelism, high memory bandwidth, and hardware architectures optimized for throughput-oriented workloads\cite{ref25,ref26}. In recent years, they have become important computing platforms for online reconstruction and fast simulation in high-energy physics. The point-to-parameter-space mapping, $\rho$-value calculation, accumulator updating, and local peak finding in the Hough transform consist of large numbers of regular and repetitive operations. Computations associated with different hits and angular samples are largely independent. The computational structure of the Hough transform is therefore well suited to GPU acceleration. Experience from experiments such as ALICE, LHCb, and CMS in GPU-accelerated track reconstruction, vertex reconstruction, event selection, and data compression further indicates that heterogeneous parallel computing has become an important development direction for HLT and fast-reconstruction systems in high-event-rate experiments\cite{ref27,ref28,ref29,ref30,ref31,ref32}.

This work investigates a GPU-parallelized Hough-transform method for track reconstruction in the STCF MDC. The method is designed to meet the HLT requirements for low latency, high throughput, and stable track reconstruction performance under high-luminosity operating conditions. Rather than directly porting the conventional serial algorithm to a GPU, the proposed method restructures the data representation and core computational operators according to the track geometry and the mathematical structure of the Hough transform. A conformal transformation is first used to map circular-arc tracks in the transverse plane into more tractable linear features, thereby reducing the complexity of the candidate-parameter search. The discrete computations performed for individual hits and angular samples are then reorganized into regular matrix-based operations. Based on the relative independence of different angular columns in the parameter space, the voting accumulation, accumulator construction, peak finding, and candidate-circle parameter extraction are decomposed into parallel tasks. The resulting data structures, memory-access patterns, and task granularity are adapted to the CUDA thread organization and GPU memory hierarchy. This design enables the parallel-computing potential of the algorithm to be more fully exploited.

The method aims to rapidly identify geometric structures consistent with genuine charged-particle tracks in high-occupancy hit data. It addresses the difficulties caused by genuine tracks being obscured by noise, weakened candidate peaks, and increased incorrect hit associations under complex background conditions. Candidate-circle parameters are geometrically associated with detector hits. This procedure removes a large number of redundant hits that are inconsistent with the candidate tracks while retaining genuine track signals as effectively as possible. It therefore provides a smaller and higher-purity input dataset for subsequent track fitting, particle-parameter estimation, and physics-event selection. The capability of the matrix-based Hough algorithm to identify candidate tracks is investigated in detail. The applicability of the GPU-accelerated Hough-transform algorithm to online MDC track reconstruction is then comprehensively evaluated. The evaluation considers genuine-signal retention under physics-background conditions, redundant-hit reduction, and end-to-end processing performance for batches of events.

This work adopts an algorithm--architecture co-design approach rather than directly porting the conventional serial algorithm to a GPU. Starting from the geometric relationships of charged-particle tracks and the mathematical structure of the Hough transform, we restructure the data representation and core computational operations. The conformal transformation maps circular-arc tracks in the transverse plane into linear features, thereby reducing the complexity of the parameter search. The discrete voting operations over individual hits and angular samples are reorganized into regular and dense matrix-based computations. This formulation exposes the inherent parallelism among the angular columns of the parameter space. Based on this formulation, the voting accumulation, accumulator construction, peak finding, and candidate-circle parameter extraction are decomposed into parallel tasks. The data layout, memory-access patterns, and task granularity are adapted to the CUDA thread organization and GPU memory hierarchy. While maintaining track reconstruction efficiency and robustness against noise, this co-design approach enables the parallel acceleration potential of the algorithm to be more fully exploited. It provides a new approach to real-time online reconstruction in high-luminosity particle-collider experiments.

\section{Algorithm Design}\label{sec:algorithm-design}

For an event to be processed, its effective hits in the STCF MDC can be represented as the point set $P = \{(x_{i},y_{i}) \mid i = 1,2,\ldots,N\}$. Here, $x_{i}$ and $y_{i}$ are the coordinates of the hits in the detector transverse plane, and N is the number of hits in the event. The trajectory of a charged particle in the transverse plane can be approximated as a circular arc\cite{ref33}. If the circle center and radius are searched directly in the original x-y plane, the parameter-space dimension is high, and the computation and storage costs will be large in events containing physics background. To simplify the candidate search for circular tracks, this work introduces the conformal transformation method, defined as:

\begin{equation}
u=\frac{x}{x^2+y^2},\qquad v=\frac{y}{x^2+y^2}.
\label{eq:conformal}
\end{equation}

Here, the point (x, y) in the x-y plane is mapped to the point (u, v) in the conformal plane after the conformal transformation. The conformal transformation preserves the local structure and direction information of the circular arc, and transforms a class of circles passing near the origin into straight lines in the conformal plane. Figure 1 shows the correspondence between circular-arc hits in the original plane and the conformal plane. This transformation converts the circular-track recognition problem into a line-recognition problem, thereby reducing the dimension of parameter search and providing a point-set representation more suitable for parallel processing in the subsequent Hough transform.

\begin{figure}[tbp]
\centering
\includegraphics[width=0.96\columnwidth]{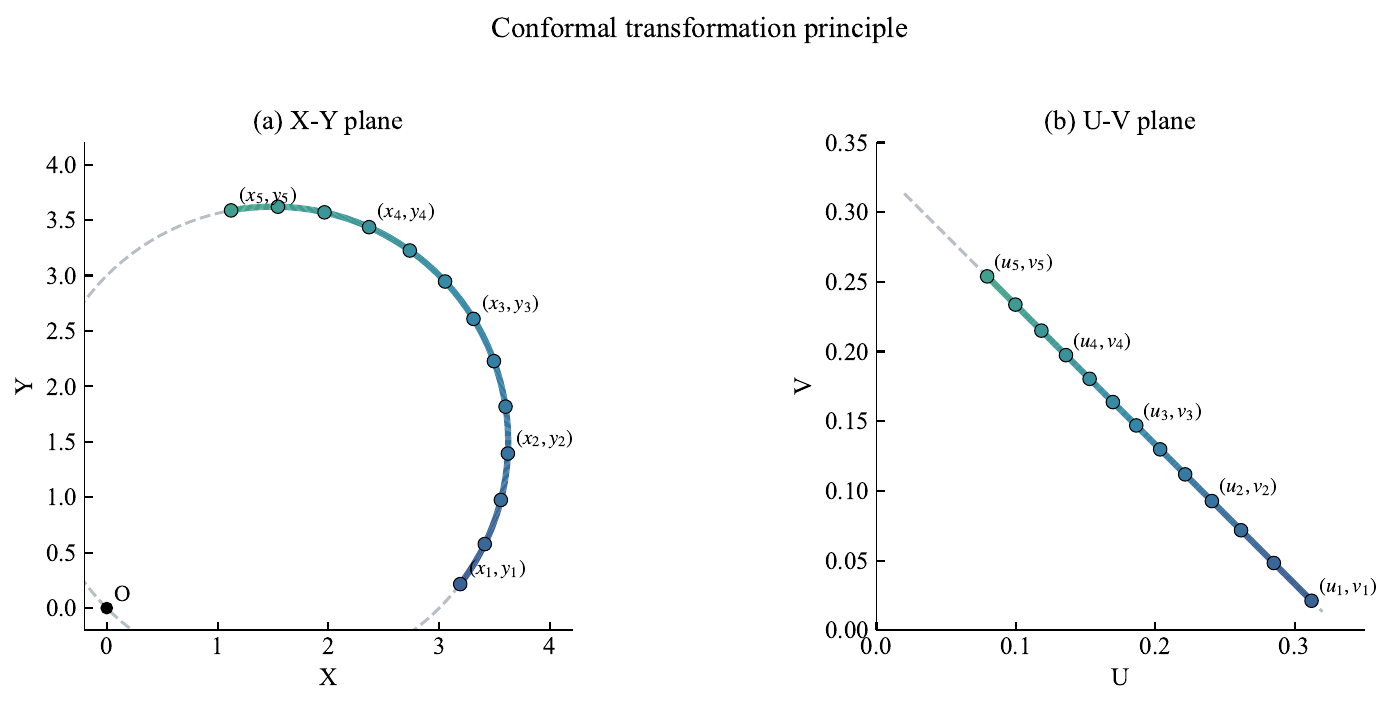}
\caption{Schematic diagram of the conformal transformation}
\label{fig:conformal}
\end{figure}

After the conformal transformation, the original problem is converted from circular-parameter search to line-parameter search. The Hough transform is used to search for line parameters from the point set in the conformal plane. For an arbitrary point $(u_{i},v_{i})$ in the conformal plane, it can be mapped by the formula $\rho = u_{i}\cos\theta + v_{i}\sin\theta$ to the $(\rho,\theta)$ parameter space. The principle of the Hough transform is shown in Fig. 2. A point in the original space corresponds to a sinusoidal curve in the $(\rho,\theta)$ parameter space. Multiple hits from the same line have corresponding curves that intersect near the same position in the parameter space, and this intersection point corresponds to a set of parameters $(\rho,\theta)$ of that line. Subsequently, candidate line parameters can be obtained by constructing a two-dimensional Hough accumulator and searching for local peaks. Using the geometric relationship of the conformal transformation, the line in conformal space can be inversely mapped to the circular-track parameters in the original x-y plane, thereby obtaining charged-particle track candidates.

\begin{figure}[tbp]
\centering
\includegraphics[width=0.96\columnwidth]{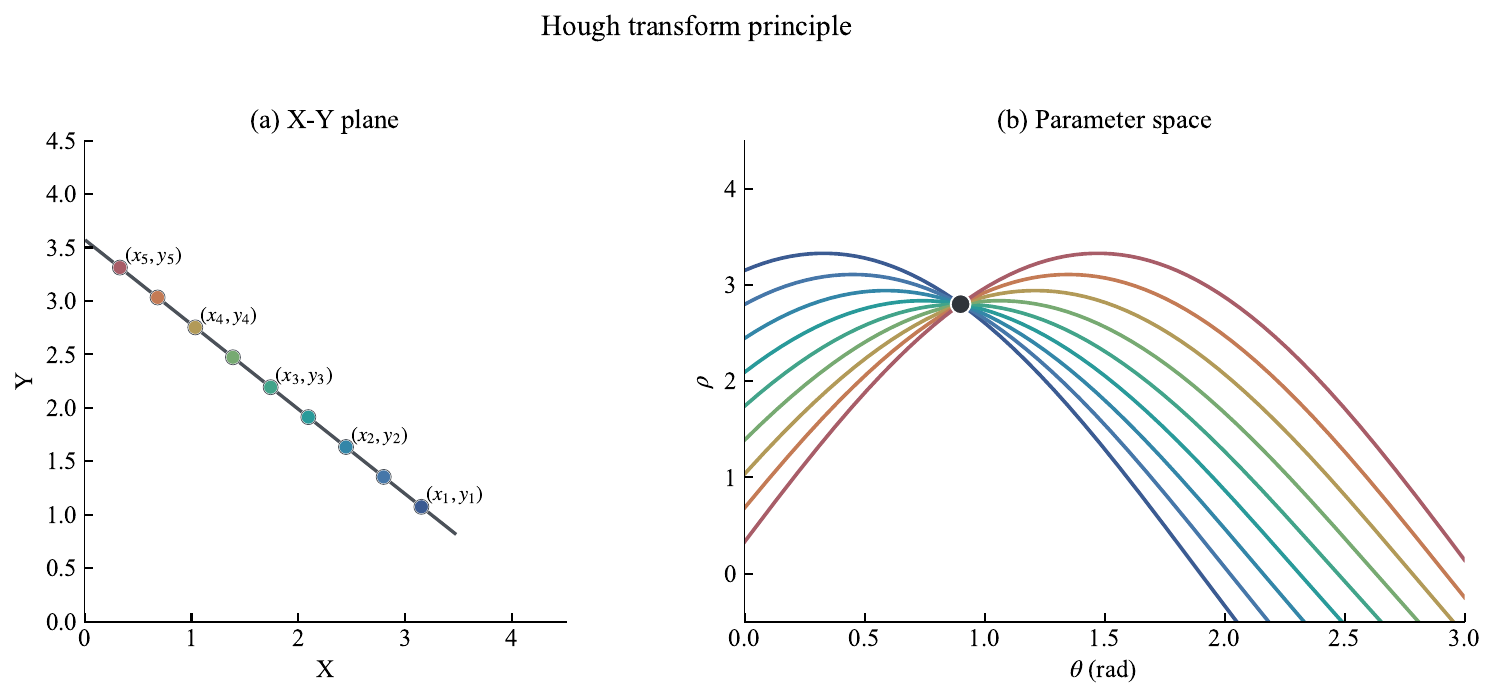}
\caption{Schematic diagram of the Hough transform}
\label{fig:hough_principle}
\end{figure}

In the Hough parameter space, the value ranges and discretization granularity of $\rho$ and $\theta$ directly affect the position and broadening of candidate peaks, thereby influencing the reconstruction error of track parameters \cite{ref17}. If the bin width is too large, different tracks with similar parameters or background hits may be accumulated in the same cell, causing peak aliasing and reducing the resolving power for candidate tracks. If the bin width is too small, hits from the same track may be dispersed into adjacent cells because of discretization errors, which weakens the voting peak \cite{ref18}. To ensure that continuous geometric parameters can be mapped stably and uniquely to the discrete accumulation space, this work discretizes the Hough parameters $\rho$ and $\theta$ at equal intervals. The range of $\rho$ is set to [-0.01, 0.01] and divided evenly into 1000 bins, corresponding to a discrete interval of $\Delta\rho = 2 \times 10^{- 5}$. The range of $\theta$ is set to $[ 0,\pi)$, and is divided evenly into 1000 bins, corresponding to a discrete interval of $\Delta\theta = \pi/1000$. This yields a parameter space of size $1000 \times 1000$, so that continuous $(\rho,\theta)$ parameters can be mapped to the corresponding discrete accumulator cells according to a uniform rule, realizing a deterministic mapping from continuous parameter representation to a regular accumulator matrix.

The local high-value regions in the two-dimensional Hough accumulation matrix reflect the common support of multiple groups of hits for the same parameter combination, and therefore track candidates can be obtained through peak finding. A relative threshold can be set according to the global maximum value of the accumulation matrix to exclude candidate peaks generated by random background and finite hit combinations. Non-maximum suppression is then performed in the neighborhood of the candidate peaks to avoid producing multiple adjacent candidates for the same physical track due to parameter discretization \cite{ref15,ref18}. For multi-track events, the algorithm adopts an iterative peak-finding strategy. After extracting the current significant peak, it suppresses its local neighborhood and continues to search for the remaining independent peaks. After the parameter set $(\rho,\theta)$ is obtained, the candidate parameter set can be inversely mapped back to the original space through Eq. (2), and data filtering can then be performed on this basis.

\begin{equation}
x_c=\frac{\cos\theta}{\rho},\qquad y_c=\frac{\sin\theta}{\rho},\qquad r_c=\frac{1}{|\rho|}.
\label{eq:circle_recovery}
\end{equation}

\section{Algorithm--Architecture Co-Design: Matrix-Based Reformulation and CUDA Implementation of the Hough Transform}\label{sec:algorithm-architecture-co-design-matrix-based-reformulation-and-cuda-implementation-of-the-hough-transform}

The conformal-Hough transform described in the previous section comprises coordinate transformation, parameter-space mapping, voting accumulation, and peak finding. These operations are regular and have weak data dependencies, which provide substantial parallelism. GPU efficiency, however, also depends on thread mapping, memory-access patterns, synchronization, and concurrent writes. The CUDA implementation should therefore reorganize the computational workflow and data layout rather than directly porting CPU nested loops to the device \cite{ref24}.

A conventional Hough transform traverses all hit-parameter combinations and updates a shared accumulator. On a GPU, this procedure can suffer from branch divergence, non-coalesced memory access, and atomic-operation contention. Concurrent updates to the same parameter bin may serialize execution. To address these limitations, the Hough mapping is reformulated as a matrix-based computation and partitioned along the angular dimension. Different angle columns can be processed independently, which reduces data dependencies and potential write conflicts. The kernels also use statically defined control flow and avoid complex object hierarchies. This design is compatible with the GPU single-instruction, multiple-thread execution model and supports coalesced memory access, shared-memory reuse, and multilevel parallelism.

Assume that one event contains $N_{p}$ effective hits and that the angular parameter is discretized into $N_{\theta}$ intervals. The conventional method evaluates every hit at every angle, giving a computational complexity of $O(N_{p} \cdot N_{\theta})$. The serial execution time therefore increases with both quantities. Different events are independent. Within an event, the calculations for different hit-angle pairs also have no recursive dependence. The implementation therefore exploits parallelism both across events and within each event.

At the event level, coordinate transformation, parameter-space accumulation, and peak extraction can be performed independently for each event. Multiple events are transferred to the GPU in batches and assigned to different thread blocks. When resources permit, they can execute concurrently on different streaming multiprocessors. Host-to-device transfer, kernel execution, and device-to-host transfer can also be overlapped. This strategy improves throughput for high-event-rate online reconstruction.

Within an event, let $(u_{i},v_{i})$ denote a hit in the conformal plane. At the discrete angle $\theta_{j}$, its Hough parameter is $\rho_{i,j}=u_{i}\cos\theta_{j}+v_{i}\sin\theta_{j}$, where $i=0,1,\ldots,N_{p}-1$ and $j=0,1,\ldots,N_{\theta}-1$. The complete set of parameter values forms the matrix

\begin{equation*}
R=
\begin{bmatrix}
\rho_{0,0} & \rho_{0,1} & \cdots & \rho_{0,N_\theta-1}\\
\rho_{1,0} & \rho_{1,1} & \cdots & \rho_{1,N_\theta-1}\\
\vdots & \vdots & \ddots & \vdots\\
\rho_{N_p-1,0} & \rho_{N_p-1,1} & \cdots & \rho_{N_p-1,N_\theta-1}
\end{bmatrix}.
\end{equation*}

Each row of \emph{R} contains the mapping of one hit over all angle samples. Each column contains the $\rho$ values of all hits at a fixed angle. The matrix is partitioned along $\theta$ into $N_{\theta}$ column vectors. The $j$th vector is [$\rho_{0,j}$, $\rho_{1,j}$, \ldots, $\rho_{N_p-1,j}$]$^T$. Different columns have no data dependence and can be assigned to separate thread blocks. Threads within a block process the hits of one column in parallel. The original nested loops are thus reorganized into event-level, angle-column-level, and hit-level parallel tasks.

\begin{figure*}[tbp]
\centering
\includegraphics[width=0.68\textwidth,height=0.78\textheight,keepaspectratio]{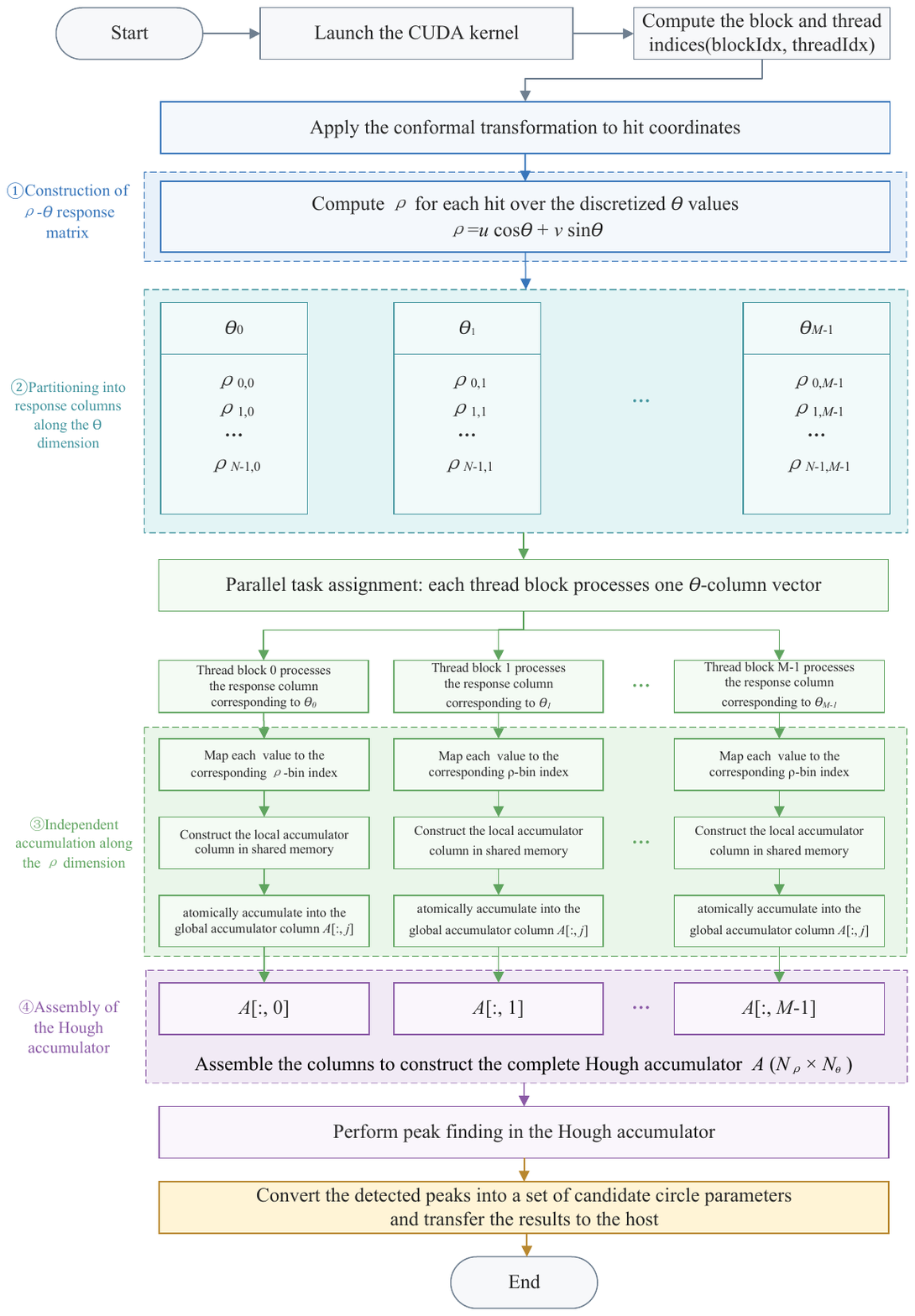}
\caption{GPU-optimized Hough-transform implementation workflow}
\label{fig:gpu_hough_workflow}
\end{figure*}

Figure 3 summarizes the implementation workflow. The parameter matrix is partitioned along the $\theta$ dimension, and each column is accumulated independently. The column results are concatenated to form the complete accumulator. This procedure converts a global accumulation into multiple local statistical tasks. Block-local accumulation reduces thread contention and the number of global atomic operations. Shared memory also reduces repeated accesses to global device memory. These optimizations improve GPU resource utilization and processing throughput. They support the real-time requirements of online track reconstruction.

Thread mapping is central to this implementation because it affects load balance, memory coalescing, synchronization, and atomic contention. Mapping one thread to each hit-angle pair exposes fine-grained parallelism, but it generates many global atomic updates. Multiple threads may also vote for the same parameter bin. Mapping one block to a hit or a group of hits improves coordinate reuse, but each block must still traverse multiple angle samples. It also does not eliminate conflicts between blocks.

The adopted strategy uses angle columns as the basic parallel units. Each thread block processes one or more $\theta$ columns. Threads within a block traverse the hit set and perform $\rho$ calculation, bin mapping, and local voting in parallel. Since the columns are independent along $\theta$, they can execute concurrently in different blocks. The dominant operations are regular multiply-add and index-mapping operations, which help limit warp divergence.

Peak finding and non-maximum suppression are also executed on the GPU. Returning the full accumulator to the host would increase device-to-host traffic and could make the CPU a bottleneck during batch processing. Only the extracted peak results need to be returned. This design reduces data-transfer overhead.

In summary, the conformal-Hough transform is implemented with event-level, angle-column-level, and hit-level parallelism. Matrix-based parameter calculation and column-wise accumulation convert serial nested loops into regular GPU tasks. Thread mapping, memory-access optimization, shared-memory reuse, and controlled atomic updates improve scalability \cite{ref19,ref20}. This implementation provides the computational basis for fast track recognition under high-event-rate conditions.

\section{Experimental Results and Analysis}\label{sec:experimental-results-and-analysis}

OSCAR is the offline software framework of the Super Tau-Charm Facility \cite{ref34}. It covers the main data-processing stages, including physics event generation, detector-response simulation, digitization, event reconstruction, and physics analysis. This framework integrates the geometry description, material information, and response models of the STCF subdetectors, and supports mixing detector hits generated by beam background with physics signal events, thereby constructing simulated data that more closely reflect actual operating conditions. Using OSCAR, physics samples with truth information can be generated under a unified software and detector model, providing reliable data basis for performance evaluation and parameter optimization of track reconstruction algorithms.

To evaluate the performance of the algorithm in a realistic experimental environment, this work uses typical representative physics-channel samples with 1x nominal background generated by OSCAR for testing. This work adopts the data compression ratio and signal retention ratio as the main evaluation metrics.

Data compression ratio is defined as the ratio of the number of candidate hits retained after Hough filtering to the total number of original hits:

\setcounter{equation}{7}
\begin{equation}
R_c=\frac{N_{\mathrm{select}}}{N_{\mathrm{total}}}.
\label{eq:dcr}
\end{equation}

where $N_{\mathrm{total}}$ denotes the total number of original detector hits in the event, and $N_{\mathrm{select}}$ denotes the number of candidate hits retained after algorithmic filtering. Under this definition, a smaller data compression ratio indicates a lower proportion of retained data and a greater reduction of redundant data. signal retention ratio is defined as the proportion of true physics hits that are successfully retained:

\begin{equation}
R_s=\frac{N_{\mathrm{true,select}}}{N_{\mathrm{true}}}.
\label{eq:srr}
\end{equation}

where $N_{\mathrm{true}}$ denotes the total number of true hits corresponding to the Monte Carlo truth track, and $N_{\mathrm{true,select}}$ denotes the number of true hits among them that are retained after filtering by the algorithm. The closer signal retention ratio is to 1, the smaller the loss of true-track information when the algorithm compresses background and redundant hits.

\begin{table*}[tbp]
\caption{Performance metrics of the Hough transform for typical physics reaction channels under onefold background}
\label{tab:dcr_srr}
\centering
\small
\begin{tabular*}{\textwidth}{@{\extracolsep{\fill}} c l c c}
\hline
Channel number & Physics analysis channel & Data compression ratio & Signal retention ratio \\
\hline
Channel 1 & \begin{tabular}[c]{@{}l@{}}$e^+e^-\to K^+K^-J/\psi$;\\ $J/\psi\to \ell^+\ell^-$\end{tabular} & 0.372 & 0.913 \\
Channel 2 & $J/\psi\to \Lambda\bar{\Lambda}$ & 0.367 & 0.836 \\
Channel 3 & $e^+e^-\to \tau^+\tau^-$ & 0.277 & 0.974 \\
Channel 4 & \begin{tabular}[c]{@{}l@{}}$e^+e^-\to \pi^+\pi^-J/\psi$;\\ $J/\psi\to e^+e^-$\end{tabular} & 0.366 & 0.961 \\
Channel 5 & \begin{tabular}[c]{@{}l@{}}$e^+e^-\to \pi^+\pi^-J/\psi$;\\ $J/\psi\to \mu^+\mu^-$\end{tabular} & 0.364 & 0.968 \\
\hline
\end{tabular*}
\end{table*}

Table 1 summarizes the performance metrics of the optimized Hough-transform algorithm for five representative physics channels, with 1,000 simulated events evaluated for each channel. The experimental data indicate that this method consistently reduces the retained hit volume across different physics processes while maintaining a high signal retention ratio. Overall, the compressed data scale is approximately 25\% to 40\% of the original data, and the signal retention ratio of most channels remains high, indicating that the algorithm mainly removes redundant hits weakly associated with target tracks and does not substantially lose signal hits during data reduction. The performance differences among different channels reflect the response characteristics of the algorithm to event topology. Among these channels, the $e^{+}e^{-} \rightarrow \tau^{+}\tau^{-}$ channel maintains good signal retention while achieving a lower data compression ratio, indicating that the algorithm has effective background-hit suppression capability for events dominated by primary-vertex tracks. The two channels containing $J/\psi$ lepton decays show similar performance, indicating that the compression performance is not sensitive to the final-state lepton type and is consistent across channels. In contrast, the signal retention ratio for the $\Lambda\bar{\Lambda}$ channel is relatively low, mainly because of secondary vertices produced by long-lived particle decays and tracks deviating from the interaction point. Such tracks have certain deviations from Hough parameterization based on primary-vertex geometric constraints. Overall, by reducing the scale of subsequent data processing, this method can still retain physics signals to a large extent, providing a basis for application to online data filtering and track preprocessing under high-event-rate conditions.

To further evaluate the track reconstruction and parameter-reconstruction performance of the Hough transform algorithm, this study uses $\mu^{+}$ single-particle samples for the analysis. Single-particle samples have clear truth information and independent track responses, which can effectively exclude the interference of background hits and multi-track overlap, and thus quantitatively test the capability of the algorithm to identify tracks.

Transverse momentum $p_{\mathrm{T}}$ is one of the core parameters characterizing the kinematic properties of charged particles in high-energy physics experiments. Under a uniform magnetic field, $p_{\mathrm{T}}$ is directly related to the curvature radius of the track in the transverse plane, and the precision of the candidate circle parameters obtained by the Hough transform directly affects the transverse-momentum reconstruction result. Therefore, the transverse-momentum residual and its distribution are important metrics for evaluating the parameter-reconstruction capability of the Hough-transform track reconstruction algorithm. This study uses the relative transverse-momentum residual $\delta p_{\mathrm{T}} = (p_{\mathrm{T}}^{\mathrm{reco}} - p_{\mathrm{T}}^{\mathrm{truth}})/p_{\mathrm{T}}^{\mathrm{truth}}$ to characterize the difference between the reconstructed value and the true value, and evaluates the systematic bias and relative resolution of transverse-momentum reconstruction through its distribution center and width, respectively. Figure 4 shows the result. The overall residual distribution is near zero, indicating that this method can maintain good agreement in transverse-momentum reconstruction in the main event samples.

\begin{figure}[tbp]
\centering
\includegraphics[width=0.96\columnwidth]{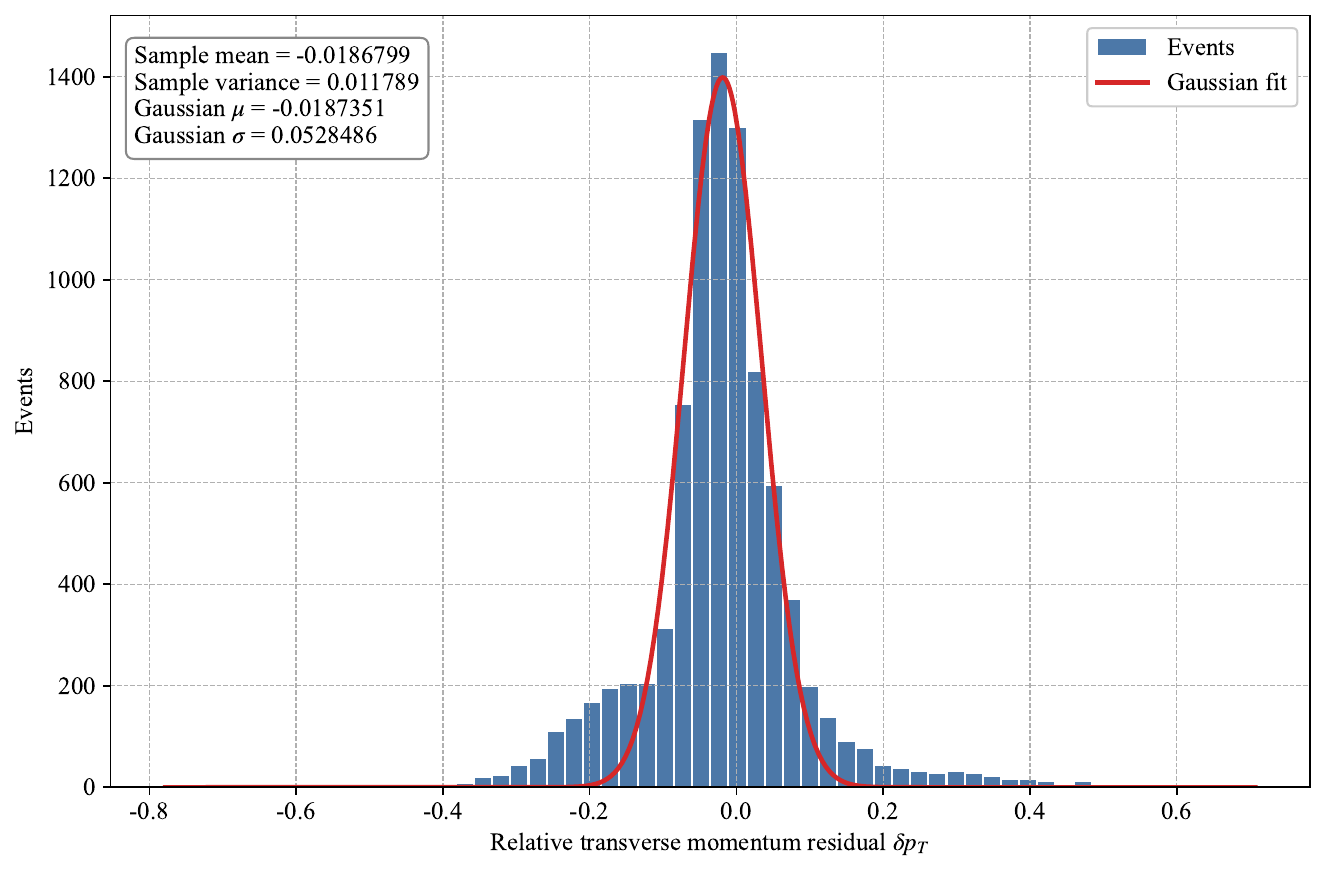}
\caption{Statistical histogram of the transverse-momentum residual distribution}
\label{fig:pt_residual}
\end{figure}

$\phi_{0}$ is an important physical quantity used in track parameterization to describe the transverse motion direction of a particle, and usually represents the azimuthal angle of the transverse momentum of a charged particle at the reference point. This reference point is generally taken as the position of closest approach of the track to the interaction point. Different from the position angle of detector hits, $\phi_{0}$ reflects the motion direction of the track in the transverse plane, and therefore directly characterizes the capability of the algorithm to reconstruct the initial emission direction of the particle. In low-transverse-momentum track reconstruction, the particle undergoes strong bending under the magnetic field, and the circle-parameter reconstruction error is further propagated to the direction-angle estimate \cite{ref34}. Therefore, $\phi_{0}$ and its residual distribution are also important metrics for evaluating the directional resolution capability of the Hough transform algorithm. This study uses $\Delta\phi_{0} = \phi_{0}^{\text{reco}} - \phi_{0}^{\text{truth}}$ to characterize the difference between the reconstructed value and the true value, and evaluates the systematic bias and angular resolution through its mean and distribution width, respectively.

Figure 5 shows the statistical distribution of $\Delta\phi_{0}$ and its Gaussian fitting result. The main body of the distribution forms a pronounced narrow peak near zero, and the fitted mean is close to zero, indicating that the reconstruction of $\phi_{0}$ by the algorithm has no significant overall systematic bias. The Gaussian width is approximately $28.8~\mathrm{mrad}$, reflecting the angular resolution level in the main event region. At the same time, the distribution has a relatively obvious non-Gaussian tail in the negative direction, shifting the sample mean toward the negative direction and causing the sample standard deviation to be significantly larger than the Gaussian fitted width. This indicates that most tracks yield stable direction-reconstruction results, but a small number of events still have large angular deviations.

\begin{figure}[tbp]
\centering
\includegraphics[width=0.96\columnwidth]{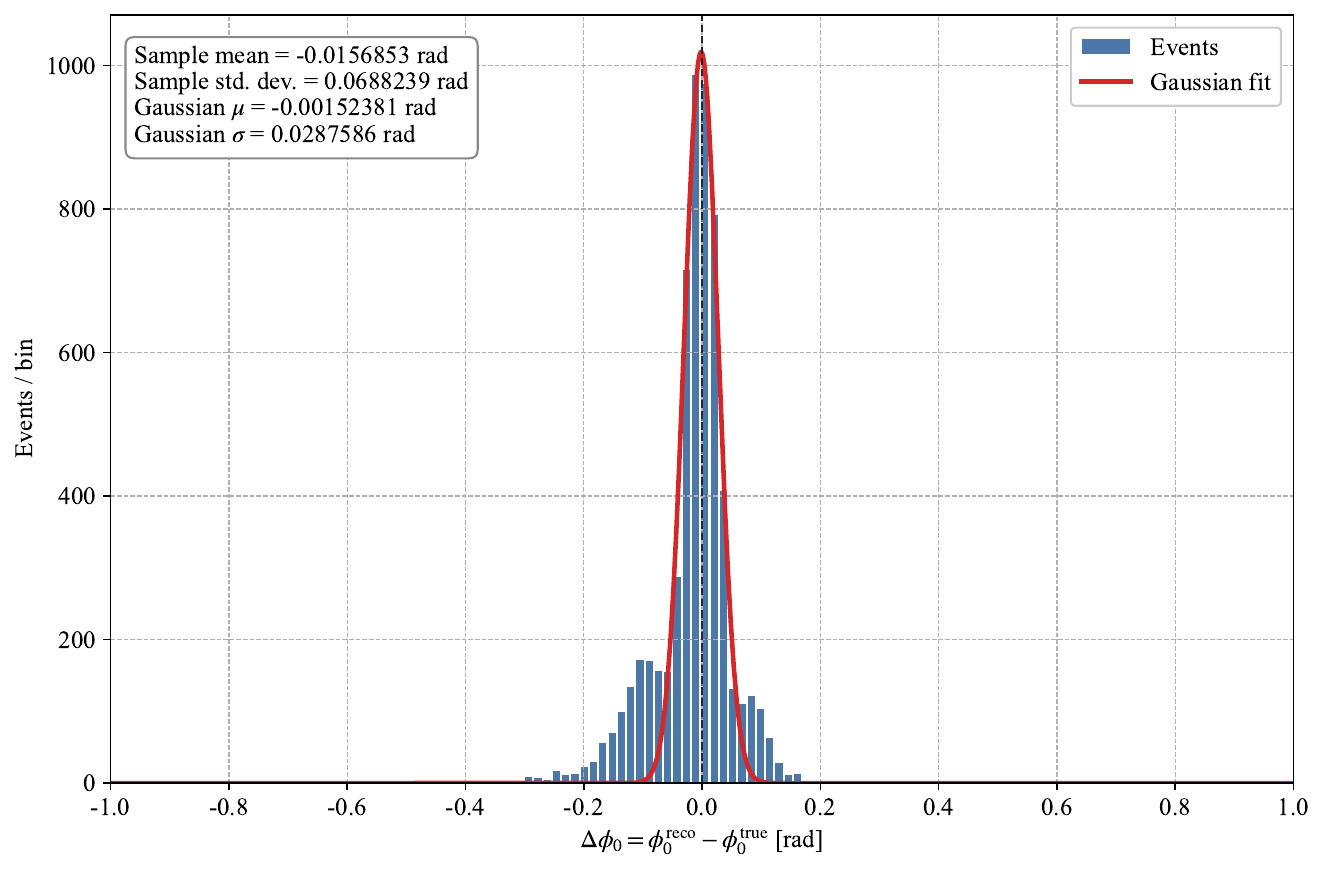}
\caption{Statistical distribution of $\Delta\phi_0$ and Gaussian fitting result}
\label{fig:dphi_hist}
\end{figure}

Based on this analysis, this work further examines the relationship between $\Delta\phi_{0}$ and particle transverse momentum, as shown in Fig. 6. The result shows that the aforementioned non-Gaussian tail in the negative direction does not originate from random fluctuations uniformly distributed over the full sample range, but is mainly attributable to the systematic negative bias of some medium- and low-transverse-momentum tracks. In this momentum interval, the particle track has a larger curvature, and the constraints of detector hits on the initial azimuthal angle are more easily affected by geometric bending and the hit distribution, which can make the reconstructed $\phi_{0}$ is systematically smaller than the true value. As the transverse momentum increases, the track curvature decreases, the binned average residual gradually returns to near zero, and the residual distribution also tends to become symmetric and stable. Therefore, the negative long tail in the overall distribution mainly originates from the bias of medium- and low-transverse-momentum tracks, rather than a global azimuthal-angle calibration bias.

\begin{figure}[tbp]
\centering
\includegraphics[width=0.96\columnwidth]{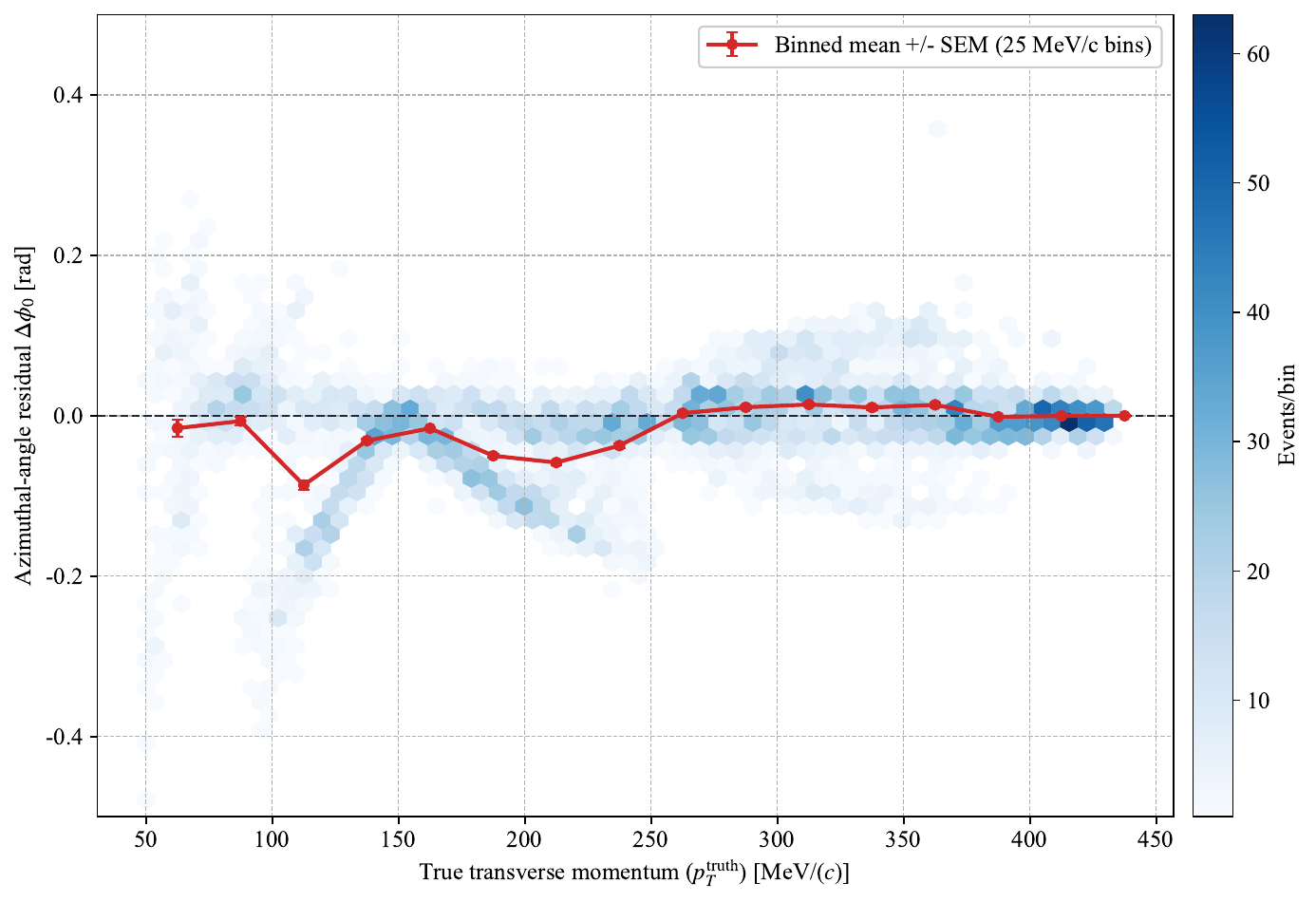}
\caption{Relationship between $\Delta\phi_0$ and particle transverse momentum}
\label{fig:dphi_vs_pt}
\end{figure}

To quantitatively evaluate the ability of the optimized Hough transform algorithm to identify tracks under different kinematic conditions, this work further studies the relationship between track reconstruction efficiency and true transverse momentum. A track is considered successfully found if it satisfies $|p_{\mathrm{T}}^{\mathrm{reco}}-p_{\mathrm{T}}^{\mathrm{truth}}|<20~\mathrm{MeV}/c$ or has a signal retention ratio greater than $90\%$. This composite criterion is adopted mainly because, when the polar angle is large, low-transverse-momentum particles may form multi-turn tracks in the detector. For such events, the main peak in the Hough parameter space corresponds only to a dominant arc segment in the track, so that the reconstructed transverse momentum is relatively consistent with the true value, but some true signal hits are not included in the candidate track, causing an apparent decrease in the signal retention ratio. On the other hand, when the candidate track has retained the vast majority of true hits, parameter-space discretization and the superposition of multi-turn arc segments may also cause a certain transverse-momentum bias. Therefore, relying on a single metric alone can easily misjudge effectively identified tracks as track reconstruction failures.

Figure 7 shows the relationship between track reconstruction efficiency and true transverse momentum in single-particle samples. The results show that the track reconstruction efficiency remains high throughout the whole test interval, with only slight fluctuations in individual transverse-momentum intervals, and shows no continuous decreasing trend with changes in track curvature. This indicates that the Hough transform has relatively stable identification capability for tracks with different nb curvatures and spatial topologies.

\begin{figure}[tbp]
\centering
\includegraphics[width=0.96\columnwidth]{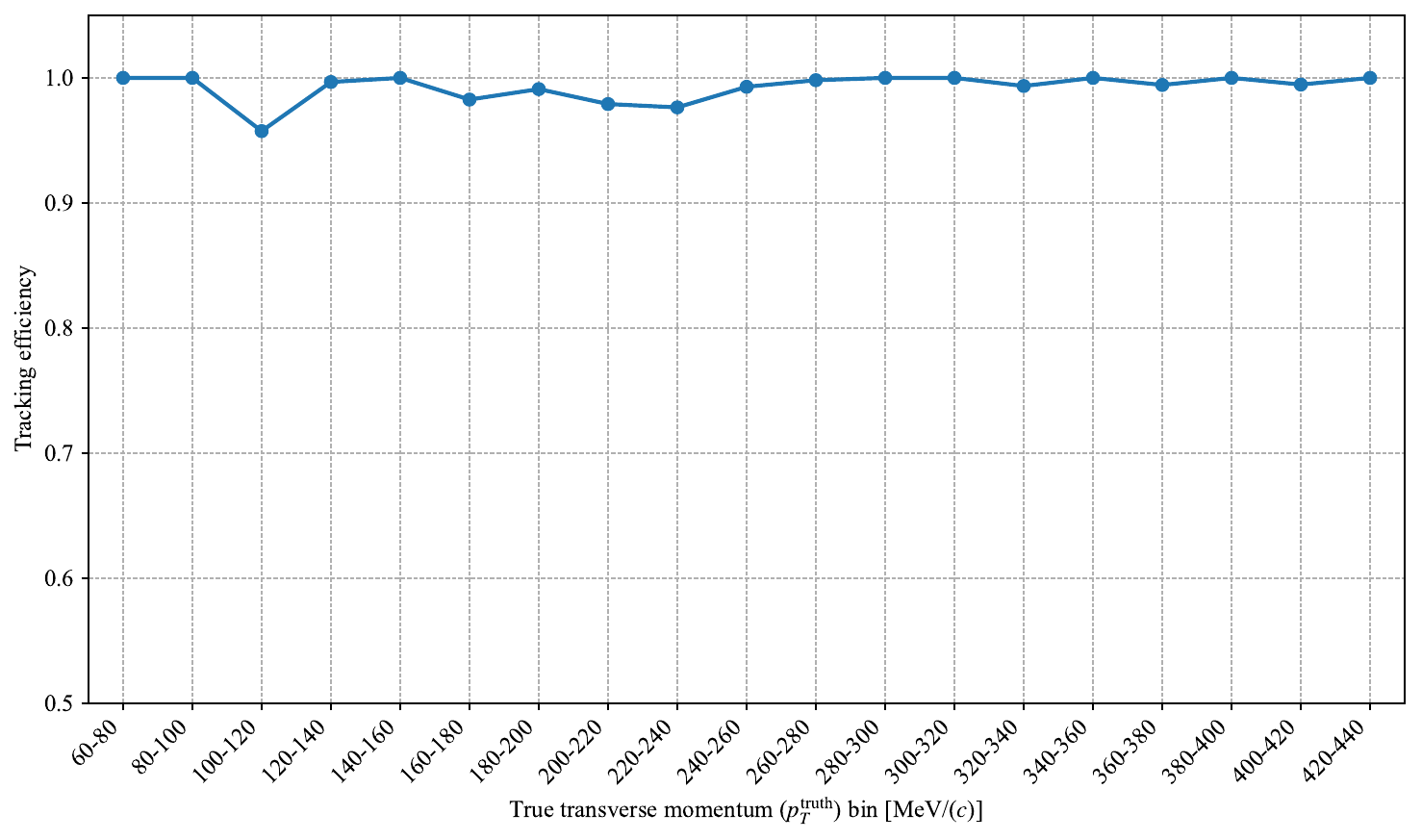}
\caption{Relationship between tracking efficiency and true transverse momentum in single-particle samples}
\label{fig:tracking_eff}
\end{figure}

After confirming physics metrics such as tracking efficiency, parameter resolution, and signal retention ratio, it is still necessary to further examine the computational performance of the algorithm in an online-processing setting. For the High-Level Trigger (HLT), track reconstruction must not only ensure sufficient physics reconstruction precision, but also process large numbers of events within a strict time budget. Otherwise, event backlogs will occur and the input event rate that the system can tolerate will be limited. Therefore, after physics performance has been established, systematic evaluation of kernel execution time, data-transfer overhead, end-to-end runtime, and acceleration relative to the CPU implementation is an important basis for judging whether the algorithm can be applied to the HLT online reconstruction workflow.

\begin{table*}[tbp]
\caption{Algorithm performance metrics on different computing platforms}
\label{tab:platform_performance}
\centering
\small
\begin{tabular*}{\textwidth}{@{\extracolsep{\fill}} c c c c c}
\hline
Implementation scheme & Computing platform & \begin{tabular}[c]{@{}c@{}}Runtime for 1000\\ events (s)\end{tabular} & \begin{tabular}[c]{@{}c@{}}Throughput\\ (events/s)\end{tabular} & Relative speedup \\
\hline
Non-matrix-based scheme & CPU & 21.22 & 47 & -- \\
\fourrowlabel{Matrix-based scheme} & CPU & 1.81 & 552 & 11.72$\times$ \\
 & RTX 5060 Ti & 0.14 & 7143 & 151.57$\times$ \\
 & RTX 4060 & 0.20 & 5000 & 106.1$\times$ \\
 & Tesla A16 & 0.35 & 2857 & 60.63$\times$ \\
\hline
\end{tabular*}
\end{table*}

Table 2 reports the end-to-end processing performance of the CPU and different GPU platforms when processing 1000 events. The test samples were selected from typical physics reaction channels $e^{+}e^{-} \rightarrow \pi^{+}\pi^{-}J/\psi$, $J/\psi \rightarrow e^{+}e^{-}$, ensuring that different implementations and computing platforms are compared under the same input data and event complexity. Taking the AMD R7 4800H as the CPU baseline platform, the conventional Hough transform takes 21.22 s to process 1000 events, whereas the matrix-based Hough implementation takes 1.81 s. In contrast, the GPU platforms markedly improve the execution efficiency of the algorithm. Under single-GPU configurations, the processing runtimes on the RTX 5060 Ti, RTX 4060, and Tesla A16 are reduced to 0.14 s, 0.20 s, and 0.35 s, respectively, corresponding to speedups of 151.57$\times$, 106.1$\times$, and 60.63$\times$. These results show that GPU platforms provide higher computational throughput than the CPU platform, mainly because the parameter-space mapping and block-wise accumulation steps in this algorithm are well suited to parallelization. These results validate the effectiveness of the proposed GPU implementation in batched-event processing and indicate its potential for online track reconstruction in trigger systems for high-energy physics.

The computational load of the Hough transform is closely related to the number of events to be processed. As the number of events in a batch increases, the total number of hits participating in parameter-space mapping grows synchronously, and the computational load of voting accumulation, peak finding, and candidate-track parameter extraction also expands. At the same time, a larger batch size also increases the pressure on intermediate-data storage, host-device data transfer, and task scheduling. Therefore, although the increase in event number does not change the basic processing workflow of a single event, it significantly amplifies the overall computational cost and imposes higher requirements on the parallel scalability of the algorithm and the efficiency of hardware-resource utilization. Based on this, this work further compares the end-to-end runtime of the CPU and RTX 5060 Ti platforms at different event scales to evaluate how the GPU adapts to increasing computational load.

\begin{figure}[tbp]
\centering
\includegraphics[width=0.96\columnwidth]{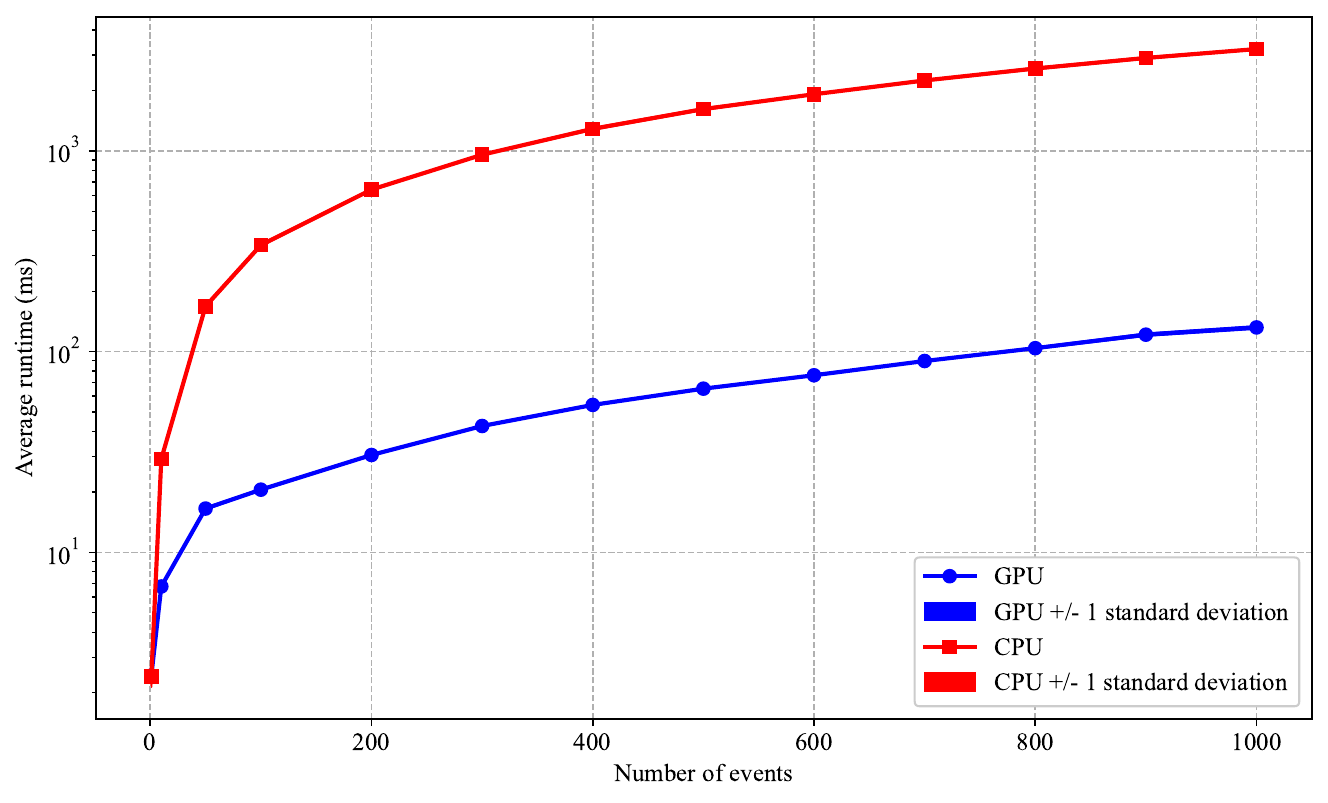}
\caption{End-to-end runtime of CPU and RTX 5060 Ti platforms at different event scales}
\label{fig:cpu_gpu_runtime}
\end{figure}

The matrix-based Hough transform algorithm was run on the CPU and RTX 5060 Ti platforms at different event scales. Each data point is the average of 10 repeated runs with the same configuration, and Fig. 8 shows the results. As the number of events increases, the CPU runtime shows clear cumulative growth, indicating that serial execution cannot fully reuse the mutually independent computing tasks among different events. In contrast, the GPU runtime grows much more slowly, and its performance advantage gradually increases as the batch-processing scale expands. This pattern indicates that, for small batches, fixed overheads such as kernel launch, data transfer, and device synchronization still account for a certain proportion of the runtime, and GPU parallel resources have not yet been fully utilized. When the batch size increases, many events can be mapped simultaneously to GPU threads for execution, the fixed overhead is effectively amortized, and the utilization of device computing units increases accordingly. The widening runtime gap between the two platforms further indicates that this algorithm has strong event-level parallelism and batch-processing scalability. The small fluctuations among repeated runs at each test point also indicate that the GPU implementation maintains stable execution performance at different task scales, and is suitable for HLT online computing scenarios with strict requirements for both throughput and processing latency.

In the GPU parallel implementation, the thread-block size directly determines the number of computing tasks carried by a single thread block, and further affects warp scheduling, computing-resource occupancy, and parallelism. For the complete GPU processing workflow constructed in this work, Hough voting, peak finding, device-memory initialization, and host-device data transfer have different computational and memory-access characteristics, so the same thread-block configuration does not have a consistent influence on all stages. To examine the effect of the thread-organization mode on end-to-end performance, this work changes the number of threads in each thread block while keeping the input data scale and algorithm parameters unchanged, and separately counts the average runtime of stages such as data transfer and device-memory initialization, Hough voting, peak finding, and result transfer back to the host, thereby analyzing the key stages that limit overall performance.

\begin{figure}[tbp]
\centering
\includegraphics[width=0.96\columnwidth]{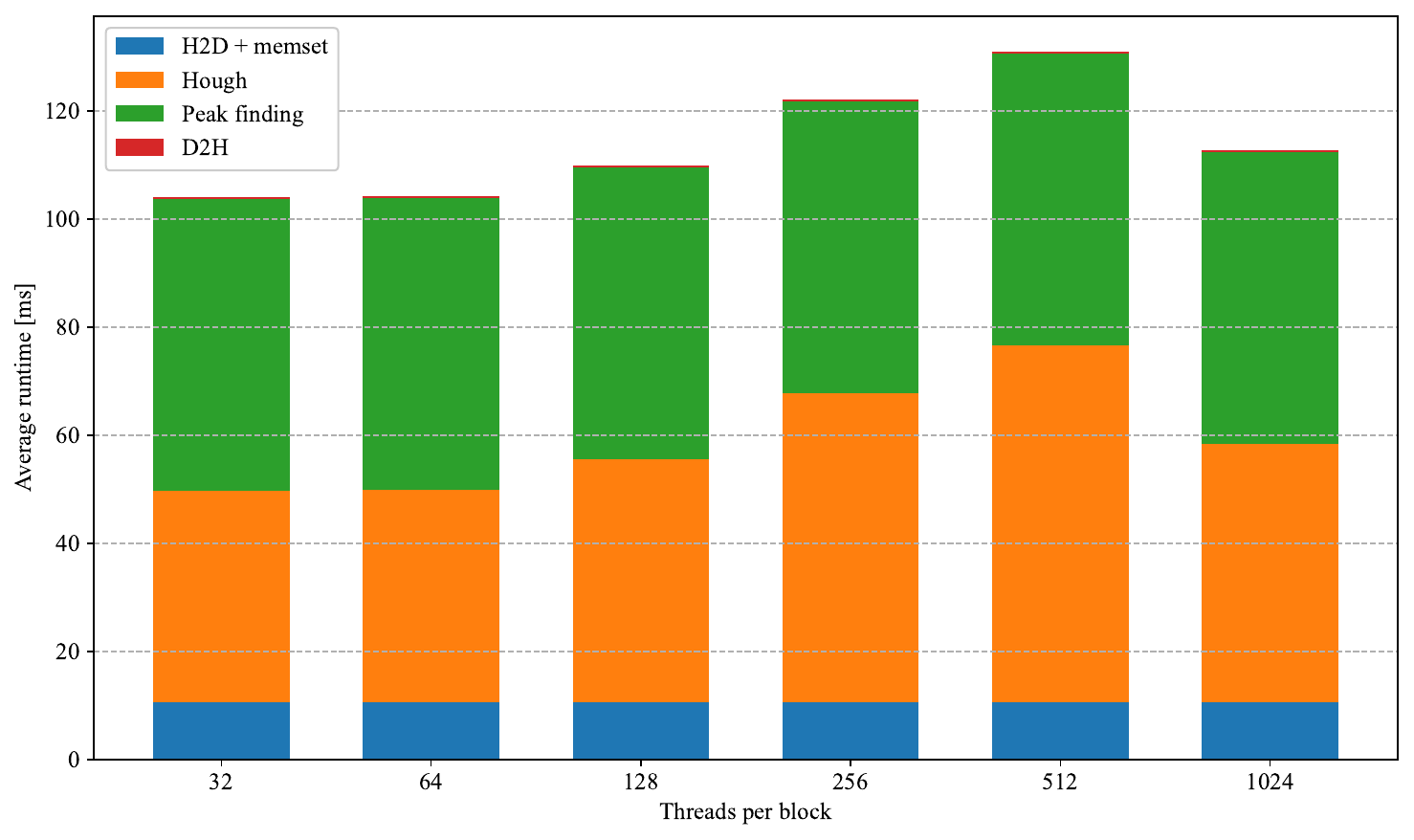}
\caption{Runtime of each GPU Hough-transform processing stage at different thread-block sizes}
\label{fig:block_size_runtime}
\end{figure}

The runtime results of each GPU Hough-transform processing stage at different thread-block sizes are shown in Fig. 9. The results show that changes in the thread-block size mainly affect the Hough voting stage, while the runtimes of data transfer, device-memory initialization, and result transfer back to the host are generally stable, indicating that these stages are not sensitive to the thread-organization mode. Although the peak finding stage accounts for a large proportion of the time, its fluctuation with the number of threads is relatively limited. Therefore, the performance difference of the complete workflow mainly originates from the Hough voting kernel. As the thread-block size increases, the total runtime does not decrease continuously, but instead increases markedly at a medium thread size, indicating that increasing the number of threads within a block does not necessarily bring higher parallel efficiency. This phenomenon may be related to a decrease in the number of resident thread blocks, competition for register and shared-memory resources, and increased thread-scheduling overhead. After the number of threads further increases, the performance of the voting stage recovers to some extent, but still does not exceed the configuration with smaller thread blocks. Thus, the thread organization of this algorithm needs to balance within-block parallelism and streaming-multiprocessor resource occupancy. Under the present test conditions, a smaller thread-block configuration achieves better end-to-end performance.

\section{Conclusion}\label{sec:conclusion}

To address the dual constraints of online track reconstruction performance and processing latency in high-event-rate high-energy physics experiments, this work proposes and implements a GPU-accelerated Hough-transform-based method for online track reconstruction in the STCF MDC, and reformulate the algorithm at two levels: parameter-space data organization and the GPU computing workflow. The conventional Hough transform performs parameter-space mapping, voting accumulation, and peak finding for a large number of detector hits. Its computational cost grows rapidly with the number of hits and the event scale, and it is difficult to directly satisfy the low-latency requirements of the High-Level Trigger system. In high-energy physics experiments, existing Hough-transform reconstruction studies mostly focus on improving the physics efficiency of low-transverse-momentum tracks or global recognition capability. The method organizes the Hough transform algorithm as matrix operations, implements the device-side execution workflow of hit mapping, accumulator construction, and candidate peak finding through CUDA thread organization, and optimizes the thread-block configuration according to the memory-access and computational characteristics of different computing stages, thereby optimizing the execution logic of the algorithm and reducing its processing runtime. This work evaluate the method in terms of physics performance, end-to-end runtime, stage overhead, batch-processing scalability, and cross-platform performance. The experimental results show that, while maintaining stable track reconstruction and parameter-reconstruction performance, this method significantly reduces the end-to-end latency of the Hough transform in the online-processing scenario: under the test condition of 1000 events per batch, the average GPU processing runtime is approximately 0.14 s, corresponding to a batch-averaged per-event processing runtime of about $0.14~\mathrm{ms}$. Compared with the CPU implementation under the same algorithmic logic, the GPU implementation achieves a speedup of $151.57 \times$ is obtained. In summary, the matrix-based organization and GPU-side optimization strategy proposed in this work effectively improve the computational efficiency of the algorithm while preserving the reliability of track reconstruction, and provide a feasible route for further development and possible integration of the Hough transform in high-event-rate HLT online track reconstruction.

The Hough transform algorithm is geometrically more suitable for processing tracks emitted from near the interaction point. For secondary-vertex particle decay tracks with large $d_{0}$, peaks in the parameter space may be weakened or dispersed into the neighborhood, thereby affecting the selection of candidate tracks. To address these problems, future work will study methods for adaptive parameter-space partitioning and dynamic peak-threshold adjustment \cite{ref17,ref18} based on concentration density and the local structure of the accumulation matrix, and will use the Hough transform as a fast candidate seed to introduce Kalman filtering for track extrapolation, hit association, and parameter-fitting correction \cite{ref35,ref36}, so as to improve the physical consistency of candidate tracks and reconstruction stability in complex-topology events. The related algorithms are under development. In terms of computational implementation, future work will explore CUDA Streams, pinned memory, and asynchronous data-transfer mechanisms suitable for the characteristics of the Hough transform algorithm, increase the overlap among data preparation, host-device communication, and kernel execution, and also explore multi-GPU task partitioning and load-balancing strategies to meet the throughput requirements for continuous streams of large event batches, improve computing-resource utilization, and continuously evaluate the reconstruction stability and online-processing capability of the algorithm under more complex background conditions.

\begin{acknowledgments}
This work is supported by the National Natural Science Foundation of China (No. 12341503); the Hunan Provincial Natural Science Foundation (No.2023RC4006); the Scientific Research Fund of Hunan Provincial Education Department (No.24B0454); the National Key R\&D Program of China (No. 2022YFA1602200 and No. 2023YFA1607200); the international partnership program of the Chinese Academy of Sciences (No. 211134KYSB20200057). We thank the Hefei Comprehensive National Science Center for their strong support on the STCF key technology research project.
\end{acknowledgments}

\end{document}